\documentclass[aps, reprint, nofootinbib]{revtex4-1}
\usepackage[sort&compress]{natbib}
\usepackage{amsmath}
\usepackage{graphicx}
\usepackage[rightcaption]{sidecap}
\usepackage{courier}
\usepackage{hyperref}
\usepackage{amsfonts}
\usepackage{color}
\usepackage{braket}

\newcommand{\be}{\begin{equation}}
\newcommand{\ee}{\end{equation}}

\pdfoutput=1
\begin{document}

\title{Upper and Lower Bounds on the Integrated Null Energy in Gravity}
\author{Stefan Leichenauer}
\email{sleichen@gmail.com}
\affiliation{Alphabet (Google) X}
\author{Adam Levine}
\email{arlevine@berkeley.edu}
\affiliation{Department of Physics, University of California, Berkeley, CA 94720, USA}
\affiliation{Kavli Institute for Theoretical Physics, University of California, Santa Barbara, CA 93106 USA}

\begin{abstract}
We prove a lower bound on the integrated null energy along achronal geodesic segments using induced gravity on a brane in AdS/CFT. The bound follows from the assumption that bulk causality respects brane causality, and matches a bound recently conjectured by Freivogel and Krommydas for semiclassical gravity. We also prove a more general upper bound on the same quantity that follows simply from achronality. We check that the lower bound is satisfied in recent constructions of traversable wormholes, and demonstrate that the bound is related to causality in the ambient spacetime of the wormhole.
\end{abstract}

\maketitle


\section{Introduction and Summary}
Many recent constraints on the energy density in quantum field theory~\cite{Bousso:2015wca,Koeller:2015qmn,Hartman:2016lgu,Faulkner:2016mzt,Balakrishnan:2017aa,Leichenauer:2018obf} were originally conjectured as statements in semiclassical gravity. In gravity, these conditions are motivated by the desire to rule out pathologies like closed timeline curves. By taking the $G_N \to 0$ limit, these bounds sometimes turn into non-trivial statements in quantum field theory, which can then be proved directly with field-theoretic techniques.

Once proven in the field theory, one can often perturbatively lift these field-theoretic statements back to semiclassical gravity. For example, the proof of the quantum null energy condition may be used perturbatively for quantum fields on a curved background, thus proving the quantum focusing conjecture, at least in certain states and limits~\cite{Bousso:2015mna}.

On the other hand, it is likely that there are additional restrictions on theories of gravity beyond those which come from quantum field theory on a curved background. Indeed, a recent conjecture by Freivogel \& Krommydas~\cite{Freivogel:2018gxj} asserts that for low energy states in a semiclassical theory of quantum gravity, there should be a semilocal bound on the null-null components of the stress tensor of the form\footnote{Outside of this introductory section we will drop explicit expectation values from the notation, but they should be understood.}
\begin{align}\label{eqn:FKBound}
\int_{-\infty}^{\infty} du\ \rho(u) \braket{T_{uu}(u)} \geq -\frac{1}{32\pi G_N} \int_{-\infty}^{\infty} du\ \frac{\rho'(u)^2}{\rho(u)},
\end{align}
where $\rho(u)$ is an arbitrary, non-negative smearing function, and the integral is over a null geodesic which is achronal on the support of $\rho$. Freivogel \& Krommydas were not able to fix the numerical factor appearing in this bound, but in this note we determine it. Notice that for a compactly-supported $\rho$, the $G_N \to 0$ limit leaves the resulting field theory energy density unconstrained. This bound also implies the achronal ANEC when applied to an inextendible achronal null geodesic, but is far more general since generic spacetimes do not possess inextendible achronal null geodescis~\cite{Graham:2007va}. This bound is also similar in flavor to the so-called quantum inequalities that have been proposed by Ford \& Roman for theories without gravity~\cite{Ford:1994bj,Ford:1996er,Ford:1999qv}.  In more than two dimensions, such a semilocal bound on the stress tensor is known to be non-existent~\cite{Fewster:aa} within field theory, so it is natural that the that the $G_N \to 0$ limit renders (\ref{eqn:FKBound}) trivial. 

In this note we prove the bound in equation~\eqref{eqn:FKBound} for holographic field theories that have been perturbatively coupled to gravity using the induced gravity framework on a brane~\cite{Randall:1999ee, Randall:1999vf, Verlinde:1999fy, Gubser:1999vj, Myers:2013lva}. The reason we use induced gravity is that all of the physics, including the low-energy gravitational physics of the brane, is encoded in the AdS dual. In particular, the induced gravity setup fixes the value for Newton's constant, as well as the higher-curvature gravitational couplings on the brane. The consistency of AdS/CFT automatically encodes certain constraints that would be impossible to guarantee if we just coupled the theory to gravity by hand. For instance, it was shown in~\cite{Myers:2013lva} that in the induced gravity setup the standard holographic entropy formula correctly computes the generalized entropy from the brane point of view, which is a nontrivial check that the induced gravity formalism encodes desirable constraints.

The main assumption in our proof of~\eqref{eqn:FKBound} is that bulk physics should respect brane causality:
\begin{itemize}
\item[]{\bf Brane Causality Condition:} The intrinsic brane causal structure cannot be violated by transmitting signals through the bulk.
\end{itemize}
In ordinary AdS/CFT (where the boundary is at infinity and not a brane at finite location), this condition was proved by Gao \& Wald~\cite{Gao:2000ga} for all asymptotically AdS spacetimes satisfying the averaged null curvature condition. However, any assumption about the bulk geometry is less fundamental than the statement of boundary causality, and one should instead use boundary causality as a basic axiom. That strategy was used in~\cite{Kelly:2014mra} to prove the ANEC for the boundary field theory and in~\cite{Levine:2016bpj} to prove the quantum inequalities. Our techniques are similar to those works, and our assumption is the Brane Causality Condition.

One may question whether the Brane Causality Condition is reasonable, even at the classical level. If the brane were an arbitrary hypersurface at finite position then surely the condition would be violated in most situations. However, the brane gravitational equations of motion save us. As we will review below, in low-energy states the brane extrinsic curvature satisfies $K_{uu}\approx 0$, so that null geodesics in the brane geometry are also null geodesics in the bulk geometry. This removes obvious violations of the Brane Causality Condition that would otherwise exist. This also highlights our earlier point about the consistency of induced gravity: coupling another matter sector to the brane metric without using induced gravity will lead to order-one violations of $K_{uu} \approx 0$, and hence of the Brane Causality Condition. In a highly curved or highly quantum regime one may question the validity of the Brane Causality Condition, but in the semiclassical regime we focus on it should be a good assumption. \footnote{Our arguments will not even make full use of the Brane Causality Condition. We only require that it be obeyed in the near-brane region of the bulk.}

As a second result, we will separately derive an upper bound on the integrated null energy in gravity, namely
\begin{align}\label{eq-upperbound}
\int_{-\infty}^{\infty} du\ \rho(u) \braket{T_{uu}(u)} \leq \frac{d-2}{32\pi G_N} \int_{-\infty}^{\infty} du\ \frac{\rho'(u)^2}{\rho(u)}.
\end{align}
Except for the factor of $d-2$, this bound is like a mirror image of~\eqref{eqn:FKBound}. In fact, this bound is much more general (and more trivial). It follows from an analogous upper bound on the integrated null curvature---obtained by multiplying~\eqref{eq-upperbound} by $8\pi G_N$ and using Einstein's equation\footnote{We freely use Einsten's equation in manipulating our inequalities even when the gravitational theory includes higher curvature terms. The assumption is that Einstein's equation is the leading part of the full gravitational equation of motion, and in low-energy states all higher-curvature terms are suppressed and therefore irrelevant for inequalities.}---that is simply a geometrical consequence of achronality. The curvature inequality holds in any spacetime, even when the spacetime is not dynamical. This is in contrast to~\eqref{eqn:FKBound}, which can be violated in an arbitrary spacetime and therefore represents an actual constraint on the states of a consistent theory of gravity.

We can summarize all of these results in the combined statement
\begin{align}
\frac{d-2}{4}\int_{-\infty}^{\infty} du \frac{\rho'^2}{\rho} \geq \int_{-\infty}^{\infty} du \,\rho R_{uu} \geq -\frac{1}{4} \int_{-\infty}^{\infty} du \frac{\rho'^2}{\rho},
\end{align}
valid for a null geodesic which is achronal over the support of $\rho$. Note that this means that, in the event that we have an inextendible achronal null geodesic, the ANCC and ANEC are actually saturated.

The remainder of this note is laid out as follows: in Section \ref{sec:II}, we will review the induced gravity formalism in the context of AdS/CFT. In Section \ref{sec:III}, we will discuss the geometric constraint imposed by brane causality. We will then use this constraint to derive~\eqref{eqn:FKBound}. Then in Section~\ref{sec-upperbound} we will derive~\eqref{eq-upperbound}, completing our main results. In Section~\ref{sec-apps} we will evaluate~\eqref{eqn:FKBound} in some recent traversable wormhole constructions which have appreciable negative energy, checking that it is not violated. Finally, in Section ~\ref{sec:discussion} we will end with a discussion of the results and possible future directions.

\section{Review of Induced Gravity on the Brane}\label{sec:II}

In this section we review some facts about the induced gravity scenarios that we will use in our computation. The construction was first used in the works of Randall and Sundrum~\cite{Randall:1999ee, Randall:1999vf}, and the relation to AdS/CFT was emphasized in~\cite{Verlinde:1999fy, Gubser:1999vj}. The extension beyond bulk Einstein gravity can be found in~\cite{Myers:2013lva}

We are interested in describing the low-energy physics of a large-$N$ field theory coupled to gravity. Because it is only an effective theory, there is an explicit UV cutoff scale. In the holographic description, this means that the asymptotically AdS space dual to the field theory has an explicit cutoff surface located at some finite position of the bulk. We will refer to this cutoff surface as the ``brane." The brane naturally has a gravitational action induced on it from the bulk gravity theory, and by ``induced gravity" we mean that, except for a few simple counterterms that we will describe below, the gravitational action for the brane consists only of the induced action from the bulk.

To aid the discussion we will introduce the coordinate $z$ normal to the brane in such a way that the metric in the vicinity of the brane is
\be\label{eq-metric}
ds^2 = \frac{dz^2 + g_{ij}(x,z)dx^idx^j}{z^2},
\ee
and the brane is located at $z=z_0$. We consider $g_{ij}({z=z_0})$ to be the physical metric of the brane. This is a rescaling of the induced metric by a factor of $z_0^2$, which is not the standard convention in induced gravity situations but is a convenient choice of units for our purposes. With this choice of metric the cutoff length scale of the effective field theory on the brane is $z_0$.


\subsection{Bulk and Boundary Actions}

The total action consists of the bulk action, a generalized Gibbons--Hawking--York brane action, and a brane counterterm action:
\be
S_\text{tot} = S_\text{bulk} + S_\text{GHY}+ S_\text{ct}.
\ee
Varying $S_\text{bulk} + S_\text{GHY}$ gives
\be\label{eq-bulkvariation}
\delta(S_\text{bulk} + S_\text{GHY}) = \int_\text{bulk} \left(\text{bulk EOM}\right) + \int_\text{brane} \sqrt{g} ~\mathcal{E}^{ij}\delta g_{ij}.
\ee
The GHY term is designed so that variation of the action only depends on $\delta g_{ij}$ and not its derivatives normal to the brane. Then we see that $\mathcal{E}_{ij}$ contributes to the brane gravitational equations of motion. For bulk Einstein gravity, $\mathcal{E}_{ij}$ is proportional to the Brown--York stress tensor, but in higher-derivative bulk gravity it will have additional terms.

The equation $\mathcal{E}_{ij} = 0$ is a higher-derivative gravitational equation of motion from the brane point of view, even when the bulk just has Einstein gravity. We will see below that, for us, it is the null-null component of this equation that matters. In the next section, when we discuss the counterterm action, we will restrict the set of allowed counterterms so that they do not affect the null-null equations of motion. The reason is that the null-null equations of motion are what ensure that the extrinsic curvature of the brane $K_{uu} \approx 0$, which is important for the Brane Causality Condition.

One important consequence of the induced gravity procedure is that the effective Newton constant on the brane is related to the bulk Newton constant by a simple rescaling:
\be\label{eq-gbrane}
G_\text{brane} = (d-2) G_\text{bulk} z_0^{d-2} + \cdots
\ee
Here the $\cdots$ refer to corrections that come from non-Einstein gravity in the bulk, but they will be suppressed by the size of the higher-curvature bulk couplings~\cite{Myers:2013lva}. We assume that those couplings are small, namely of the order typically generated by bulk quantum effects. Since we are interested in proving inequalities like~\eqref{eqn:FKBound}, only the leading-order parts of our expressions are important, and so terms like this can be dropped.

We would also like to emphasize that the construction of the brane theory is identical to the first few steps of the standard holographic renormalization procedure~\cite{deHaro:2000vlm}. In holographic renormalization, one would introduce counterterms that cancel out the purely geometric parts of $\mathcal{E}_{ij}$, and the part that remains is the holographic stress tensor. Here we do not introduce most of those counterterms (the exceptions are described below), and instead interpret those purely geometric parts of $\mathcal{E}_{ij}$ as the geometric terms in the gravitational equations of motion. The upshot is that the ordinary holographic stress tensor still has the same interpretation in the induced gravity scenario as it does in ordinary AdS/CFT: it is the stress tensor of the matter sector of the theory, and it plays the role of the source in the gravitational equations of motion.


\subsection{Counterterms}

Now we discuss the counterterm action, $S_\text{ct}$. The purpose of the counterterm action is to fine-tune the values of certain mass parameters in the induced theory which would otherwise be at the cutoff scale. This includes the brane cosmological constant, which can be tuned by adding a term to $S_\text{ct}$ of the form
\be\label{eq-braneaction1}
S_\text{ct} \supset \int_\text{brane} \sqrt{g}\ \mathcal{T},
\ee
where the constant $\mathcal{T}$ is known as the tension of the brane.

No other purely gravitational counterterms will be added to the brane action. As mentioned in the introduction, the fact that the brane gravity is induced by the bulk gravity is an important constraint that enforces consistency conditions which are not apparent from the effective field theory point of view. A counterterm proportional to the Einstein--Hilbert action, for example, would change the value of the brane Newton constant away from~\eqref{eq-gbrane}, and thus take us out of induced gravity. From a more practical point of view, we discussed above that the Brane Causality Condition is sensible because $K_{uu} \approx 0$, and that is enforced by the null-null equation of motion determined by $S_\text{bulk} + S_\text{GHY}$. To preserve that condition we need that $S_\text{ct}$ has a trivial variation with respect to the null-null components of the metric. This is true for the cosmological constant counterterm, and in fact is true for any counterterm that only depends on the metric through the volume element $\sqrt{g}$.

When there are low-dimension scalar operators in the field theory, new counterterms are needed to fine-tune their masses and expectation values. These include terms proportional to $\int_\text{brane} \sqrt{g} \Phi^2$, where $\Phi$ is the bulk field dual to the operator, familiar from the theory of holographic renormalization. Like the cosmological constant term, these only depend on the metric through $\sqrt{g}$, and so we can add them freely.  We will not say any more about these kinds of terms, as they are not important for the rest of our analysis.


\subsection{Brane Equations of Motion}

Now that we have discussed the action for the induced gravity system, we can calculate the correct gravitational equation of motion. Since all of the terms in $S_\text{ct}$ are coupled to the metric through $\sqrt{g}$, the result is simple. We find
\be
\mathcal{E}_{ij} \propto g_{ij},
\ee
where the proportionality factor could depend on scalar expectation values.

For Einstein gravity in the bulk, this equation sets the extrinsic curvature to be proportional to the metric:
\begin{align}\label{eqn:braneeq}
K_{ij} \propto g_{ij},
\end{align}
where
\begin{align}\label{eqn:extcurv}
K_{ij} = \frac{1}{2z}\partial_z g_{ij} - \frac{1}{z^2} g_{ij}.
\end{align}
Note that the null-null components of the extrinsic curvature would be set to zero according to this equation. For higher-derivative bulk gravity there will be corrections that we comment on below.

When written in terms of brane quantities, the equation of motion takes the form of Einstein's equation plus corrections:
\be
R_{ij} = 8\pi G_\text{brane} T_{ij} + \cdots.
\ee
The higher-curvature terms in $\cdots$ are suppressed by the brane cutoff scale, and so can be consistently dropped in states where the brane curvature scale is well below the brane cutoff scale.

Finally, we quote one additional fact which follows from standard Gauss--Codazzi-like relations on the brane, and that is the following expression for the normal derivative of the extrinsic curvature:
\begin{align}\label{eqn:Ricci}
z\partial_z K_{ij} = R_{ij} - \mathcal{R}_{ij} - z^2K K_{ij} + 2z^2K_{ik}K^k_j,
\end{align}
where $R_{ij}$ and $\mathcal{R}_{ij}$ are the brane and bulk Ricci tensor, respectively.\footnote{Note that we are raising indices in this equation using $g_{ij}$, not $\gamma_{ij} = g_{ij}/z^2$.} Together with the brane equation of motion, this equation will allow us to prove~\eqref{eqn:FKBound} in the next section.


\section{Lower Bound from Brane Causality}\label{sec:III}
In this section, we derive the bound in~\eqref{eqn:FKBound} from the Brane Causality Conditon. The technique is very similar to that used to derive the ANEC~\cite{Kelly:2014mra} and quantum inequalities~\cite{Levine:2016bpj} in AdS/CFT, with the main difference being that the brane is at a finite location in the bulk, rather than at infinity, and its intrinsic and extrinsic geometry are determined by equations of motion.

Consider a future-directed achronal null geodesic segment on the brane (defined according to the brane metric), parametrized by affine parameter $\lambda$ that takes values in the range $\lambda_0 <\lambda < \lambda_1$. We will define the null coordinate $u$ such that $u = \lambda$, and let $v$ be another null coordinate in the neighborhood of the geodesic such that $v=0$ and $g_{uv} = -1$ along the geodesic itself. We extend these coordinates into the bulk in an arbitrary way, provided that they remain orthogonal to the $z$ coordinate so that~\eqref{eq-metric} is respected. The Brane Causality Condition states that any future-directed causal curve anchored to the brane---including those which travel through the bulk---beginning at $(u,v) = (\lambda_0,0)$ must have its other endpoint in the future of our null geodesic segment according to the causal structure of the brane metric.

To derive~\eqref{eqn:FKBound}, we will construct a causal curve which begins at $(u,v) = (\lambda_0,0)$ on the brane and travels through the bulk before returning to the brane. The restriction that the curve is causal means that
\be\label{eq-bulkcausal}
\left(\frac{dZ}{d\lambda}\right)^2 + g_{ij}(X,Z)\frac{dX^i}{d\lambda}\frac{dX^j}{d\lambda} \leq 0,
\ee
where $X^i(\lambda)$ and $Z(\lambda)$ are the coordinates of the bulk curve.

To get the strictest bound, we will try to construct a bulk curve which moves as quickly as possible while remaining causal (i.e., gets infinitesimally close to being null in the bulk). Thus, we choose the bulk curve to follow a trajectory very close to the geodesic segment on the brane:
\begin{align}
z &= Z(\lambda) = z_0+ \epsilon \sqrt{\rho(\lambda)},\\
u &= U(\lambda) = \lambda,\\
v&= \epsilon^2 V(\lambda).
\end{align}
The function $\rho$ is non-negative, smooth, and satisfies $\rho(\lambda_0) = \rho(\lambda_1) = 0$, but is otherwise arbitrary. Here $\epsilon$ is a small length scale, and we should say how small it is relative to the other scales in the problem. Recall that the cutoff scale for the brane theory is $z_0$, and let us denote the characteristic curvature scale on the brane in the state we consider by $\ell$. Then we want our parameters to be such that
\be
z_0 \ll \epsilon \ll \ell.
\ee
The idea here is that our bulk curve is not probing the deep UV of the theory, where quantum gravity effects may become large, but is still microscopic compared to the curvatures scales of the state we are in. The fact that $\ell \gg z_0$ is part of the semiclassical assumption.

Expanding~\eqref{eq-bulkcausal} in $\epsilon$ out to $O(\epsilon^2)$, we find
\begin{align}\label{eqn:causality}
\epsilon \sqrt{\rho}\partial_zg_{uu}+\epsilon^2 \left(\frac{1}{4}\frac{\rho'^2}{\rho} + \frac{1}{2}\rho \partial^2_zg_{uu}-2V'\right)\leq 0.
\end{align}
All metric factors are being evaluated at $z=z_0$ along the null geodesic segment. 

Consider the $O(\epsilon)$ term. If the bulk theory were pure Einstein gravity, then from~\eqref{eqn:braneeq} and~\eqref{eqn:extcurv} we would have $\partial_zg_{uu} = 0$ on the brane. This would be violated by a small amount in higher-curvature bulk theories. Even in that case, we know from the Fefferman-Graham expansion of the metric that, generally, $\partial_zg_{uu}\propto z_0$~\cite{Imbimbo:1999bj, Schwimmer:2008yh}. Thus the $O(\epsilon)$ term is negligible for multiple reasons compared to the $O(\epsilon^2)$ term, and so we can consistently drop it from the inequality.

For the $O(\epsilon^2)$ term, the main problem is evaluating $\partial^2_z g_{uu}$ on the brane. This is easily accomplished using \eqref{eqn:Ricci}, along with the brane equations of motion. In the case of bulk Einstein gravity, from~\eqref{eqn:braneeq} we find that
\be
\frac{1}{2}\partial_z^2g_{uu} = R_{uu} -\mathcal{R}_{uu}.
\ee
For non-Einstein gravity in the bulk, there will be small corrections to this equation proportional to the bulk curvature couplings. But since those couplings are small, all of those correction terms can be dropped while preserving the inequality.

We find that~\eqref{eq-bulkcausal} reduces to
\be
\frac{1}{4}\frac{\rho'^2}{\rho}  +\rho \left(R_{uu}-\mathcal{R}_{uu}\right)- 2V'  \leq 0.
\ee
We can satisfy this condition by choosing
\be
V(\lambda) = \frac{1}{2}\int_{\lambda_0}^\lambda \rho \left(R_{uu}-\mathcal{R}_{uu}\right)d\tilde\lambda + \frac{1+\delta}{8}\int_{\lambda_0}^\lambda \frac{\rho'^2}{\rho} d\tilde\lambda
\ee
Here $\delta >0$ is a regulator that we will eventually take to zero. Thus the total change in the $v$ coordinate over the entire trajectory is
\be
\Delta v = \epsilon^2\left(\frac{1}{2}\int_{\lambda_0}^{\lambda_1} \rho \left(R_{uu}-\mathcal{R}_{uu}\right)d\tilde\lambda + \frac{1+\delta}{8}\int_{\lambda_0}^{\lambda_1} \frac{\rho'^2}{\rho} d\tilde\lambda\right).
\ee
Now we impose the Brane Causality Condition, which demands that $\Delta v \geq 0$. This must be true for any $\delta$, so in the limit $\delta \to 0$ we find the inequality
\begin{align}\label{eqn:curvbound}
\int_{\lambda_0}^{\lambda_1} \rho \left(R_{uu}-\mathcal{R}_{uu}\right)d\tilde\lambda \geq -\frac{1}{4}\int_{\lambda_0}^{\lambda_1} \frac{\rho'^2}{\rho} d\tilde\lambda.
\end{align}
We are free to formally let $\lambda_0 \to -\infty$ and $\lambda_1 \to +\infty$ as long as the geodesic is achronal on the support of $\rho$.

Now we will argue that $\mathcal{R}_{uu}$ should be dropped from the inequality, which will complete the proof. From the bulk equations of motion, $\mathcal{R}_{uu} \approx 8\pi G_\text{bulk}T^\text{bulk}_{uu}$. When written in terms of expectation values of operators in the brane field theory the slowest possible falloff at small $z_0$ is $T^\text{bulk}_{uu} \propto z_0^{2\Delta}$ with $2\Delta > d-2$ by the unitarity bound. On the other hand, $R_{uu} \approx 8\pi G_\text{brane} T_{uu}$ and $G_\text{brane} \sim z_0^{d-2}$ from~\eqref{eq-gbrane}. Thus at small $z_0$ the $\mathcal{R}_{uu}$ term is negligible, and we recover~\eqref{eqn:FKBound}.


\section{Upper Bound From Achronality}\label{sec-upperbound}

In this section, we note that achronality actually also implies an upper bound on the null curvature. This bound will be purely geometric and apply equally well to dynamical and non-dynamical backgrounds, though in theories of gravity we can turn it into the bound~\eqref{eq-upperbound} on the null energy density.

The setup is the same as before, where we have a future-directed achronal null geodesic segment with affine parameter $\lambda$ such that $\lambda_0 < \lambda < \lambda_1$. Choose some smooth function $\rho(\lambda)$ such that $\rho(\lambda_0) = \rho(\lambda_1) = 0$. We will assume that $\lambda_0$ and $\lambda_1$ are both finite at first, and we will allow them to go to infinity later as part of a limiting procedure. Then we can perform the Weyl transformation
\begin{align}
g_{ij} \to \tilde{g}_{ij} = \rho^{-1} g_{ij}
\end{align} 
in a neighborhood of the segment (after choosing some suitable extension of the affine parameter to that neighborhood). Since causal structure is preserved by Weyl transformations, in the new spacetime our segment is actually an inextendible achronal null geodesic. Note that $\lambda$ no longer affinely-parameterizes the geodesic, but we can pick a new affine parameter $\tilde \lambda$ defined by the generator $\tilde{k}^i= \left(\partial_{\tilde\lambda}\right)^i = \rho k^i$, where $k^i = \left(\partial_\lambda\right)^i$ is the generator in the original spacetime. The endpoints of the geodesic are at $\tilde\lambda = \pm\infty$, which confirms that the geodesic is inextendible.

A key fact is that the conformal transformation properties of the Ricci curvature imply that
\be\label{eq-weyl}
\int_{\lambda_0}^{\lambda_1}d\lambda \left(\rho R_{ij}k^ik^j - \frac{d-2}{4}\frac{\rho'^2}{\rho}\right) = \int_{-\infty}^{\infty} d\tilde{\lambda}\ \tilde{R}_{ij} \tilde{k}^i \tilde{k}^j.
\ee
Thus to prove~\eqref{eq-upperbound} we only have to show that the integrated null curvature on the right-hand-side is negative. Since we are assuming $\lambda_0$ and $\lambda_1$ are finite---and that the curvature in the original spacetime does not have singularities---we see from the expression on the left-hand-side that the integrated null curvature in the new spacetime is bounded.

Since our geodesic is inextendible and achronal in the new spacetime, it must be that a null congruence starting at $\tilde\lambda = -\infty$ with vanishing expansion (and twist) does not encounter a caustic at any point along the geodesic, meaning that the expansion $\tilde\theta$ remains finite as a function of $\tilde\lambda$. Integrating Raychaudhuri's equation gives
\begin{align}
\tilde\theta(+\infty) = \int_{-\infty}^\infty d\tilde\lambda\left( -\frac{\tilde{\theta}^2}{d-2} - \tilde{\sigma}^2 - \tilde{R}_{ij} \tilde{k}^i \tilde{k}^j\right).
\end{align}
If the integrated null curvature is positive, then $\tilde\theta(+\infty)$ is negative. But then the integral of $\tilde\theta^2$ diverges and we learn that actually $\tilde\theta(+\infty)$ itself is divergent. By making the same argument at large-but-finite $\tilde\lambda$, we can also rule out the possibility that $\tilde\theta$ oscillates between positive and negative values as it diverges. We will now show that $\tilde\theta$ cannot diverge at infinity, which proves the result.

Under the assumption that $\tilde\theta$ diverges at infinity, consider dividing Raychaudhuri's equation by $\tilde\theta^2$ first and then integrating from some $\tilde\lambda_0$ to $\tilde\lambda_1$, with $\tilde \lambda_0$ chosen large enough so that $\tilde\theta$ does not vanish for any $\tilde \lambda > \tilde\lambda_0$. We find
\be
\frac{1}{\tilde\theta_0} - \frac{1}{\tilde\theta_1} + \int_{\tilde\lambda_0}^{\tilde\lambda_1} d\tilde\lambda\ \frac{\tilde{R}_{ij} \tilde{k}^i \tilde{k}^j}{\tilde\theta^2} = -\frac{\tilde\lambda_1-\tilde\lambda_0}{d-2} - \int_{\tilde\lambda_0}^{\tilde\lambda_1}d\tilde\lambda\ \frac{\tilde\sigma^2}{\tilde\theta^2}.
\ee
Given the finiteness of the integrated null curvature, we see that the left-hand-side of this equation goes to a constant as $\tilde\lambda_1 \to \infty$ while the right-hand-side diverges. Thus we have proved the inconsistency of $\tilde\theta$ diverging at infinity, and the desired result follows.


\section{Applications}\label{sec-apps}

We now discuss possible applications of this bound to semiclassical gravity. In the regime of weak gravity, we might worry that the bound is trivial because $1/G_N$ is large compared to the size of the stress tensor. However, we can make up for this if the geodesic is long enough. Clearly in the case of an infinite geodesic the bound~\eqref{eqn:FKBound} implies the achronal ANEC, which is not a trivial statement. For finite but long geodesics we can get relatively strong lower bounds by choosing $\rho$ to slowly ramp up from zero to one, say by choosing $\rho = (\lambda-\lambda_0)^2/(\Delta\lambda)^2$ for some interval $\lambda_0 < \lambda < \lambda_0 + \Delta \lambda$, before transitioning to $\rho=1$. Then the integral of $\rho'^2/\rho$ is of order $1/\Delta \lambda$. Thus if $\Delta \lambda \sim 1/G_N$ we can get $O(G_N^0)$ lower bounds on the integrated null energy, assuming that most of the null energy flux is in the part of the geodesic where $\rho =1$.

In the remainder of this section we will apply the above strategy to two recent constructions of traversable wormhole solutions, which make critical use of negative energy. We will see how the achronality condition prevents each from violating~\eqref{eqn:FKBound}.


\subsection{Gao--Jafferis--Wall Wormhole}

In~\cite{Gao:2016bin} a wormhole in the bulk is made traversable by coupling two holographic CFTs in the thermofield double state. The coupling breaks achronality of the black hole horizon, thereby allowing negative averaged null energy along the horizon without violating the achronal ANEC. However,~\eqref{eqn:FKBound} still applies, and we can see what consequences it has. This is a case where the stress tensor is perturbative and $O(N^0)$ in large-$N$ counting, while the lower bound is $O(N^2)$. One might hope that applying the strategy above to reduce the magnitude of the lower bound would help here, but it does not: one can check that in situations where the geodesic becomes long enough to appreciably decrease the magnitude of the lower bound, the magnitude of the integrated energy flux decreases by an even larger factor.\footnote{We thank Don Marolf for discussions on this point.} Thus the bound never becomes tight for this construction.

\subsection{Maldacena--Milekhin--Popov Wormhole}

In~\cite{Maldacena:2018gjk} the authors constructed a traversable wormhole in four-dimensional asymptotically flat space threaded by magnetic flux and supported by the negative Casimir energy of a fermion field. The wormhole interior is given by an approximate $AdS_2 \times S_2$ metric,
\be
ds^2 \approx r_e^2\left(-(1+\xi^2)\frac{dt^2}{\ell^2} + \frac{d\xi^2}{1+\xi^2} + d\Omega_2^2\right),
\ee
where $r_e$ parameterizes the size of the wormhole and $\ell$ is such that the $t$ coordinate smoothly maps onto the Minkowski $t$ coordinate outside the wormhole. This metric is only a good description for $|\xi| \lesssim \xi_c \sim \ell/r_e \gg 1$, where it opens up into the asymptotically flat ambient space. 

We can use $\xi$ as the affine parameter of a null geodesic that passes through the wormhole, and we need to integrate the null Casimir energy along the geodesic. From solving Einstein's equations, one learns that there is a relationship between the energy density and the parameter $\xi_c$. The end result is $\ell^2T_{tt} = (1+\xi^2)^2T_{\xi\xi} \sim -1/G_N\xi_c$, which means that the integrated null energy is
\be
\int d\xi\ \left(T_{\xi\xi} + \ell^2\frac{T_{tt}}{(1+\xi^2)^2}\right) \sim -\frac{1}{G_N\xi_c}~,
\ee
with most of the contribution coming from the region $\xi \lesssim 1$.

Naively, one would consider a geodesic which went through the entire wormhole, $-\xi_c < \xi < \xi_c$, and by appropriately choosing $\rho(\xi)$ one could make $\int \rho'^2/\rho \sim 1/\xi_c$. In that case we would parametrically saturate~\eqref{eqn:FKBound}, and it would be up to the order-one coefficients to determine if the bound were in danger of being violated. However, this is too fast and we first need to properly account for the achronality condition.

In the ambient flat space, the two ends of the wormhole are a proper distance $d$ apart, which means it takes a time $d$ to send a signal from one to the other. Sending a signal through the wormhole would take a time
\be
\int_{-\xi_c}^{\xi_c} \frac{\ell d\xi }{1+ \xi^2} \approx \pi \ell,
\ee
which one expects to be greater than $d$ so that the wormhole respects the ambient causality. Define $y = \pi \ell / d$. In the solutions of~\cite{Maldacena:2018gjk} the minimal value of $y$ was approximately $2.35$, and $y =1$ means that ambient causality is being saturated.

If $y>1$ then it is faster to travel through the ambient space than it is through the wormhole, and so the null geodesic which passes through the entire wormhole from end to end is not achronal. In order to maintain achronality, we need to restrict the null geodesic segment to lie within the range $|\xi | < \xi_1$ where
\be
\arctan \xi_1 = \frac{\pi}{4}\left(1+ \frac{1}{y}\right) - \frac{1}{2\xi_c},
\ee
in the approximation that $\xi_c \gg 1$. We see that when $y=1$ we have $\xi_1 \sim \xi_c$ and~\eqref{eqn:FKBound} would be parametrically saturated, if not violated. However, if $y$ is appreciably larger than $1$, as it is in~\cite{Maldacena:2018gjk}, then $\xi_1 \sim 1$ and we are far from saturating~\eqref{eqn:FKBound}. Thus it seems that~\eqref{eqn:FKBound} is intimately connected with causality in the ambient space.


\section{Discussion and Future Directions}\label{sec:discussion}

The obvious next goal would be to prove~\eqref{eqn:FKBound} without using induced gravity. Our method of proof involved an extension of bulk-boundary causality to the brane at $z=z_0$. This suggests that the bound (\ref{eqn:FKBound}) is to be related to some notion of causality in the gravitational theory. In~\cite{Akers:2016aa}, it was shown that the analogous condition in ordinary AdS/CFT was implied by the principle of entanglement wedge nesting. Furthermore, in \cite{Balakrishnan:2017aa} it was shown that entanglement wedge nesting can be re-cast as a statement of causality under modular time evolution. It would be interesting to understand if (\ref{eqn:FKBound}) is related to some notion of modular causality in effective gravitational theories. An investigation along these lines would also have to confront the fact that the naive generalization of entanglement wedge nesting to the brane case is almost always violated.

Recently, the bound of \cite{Engelhardt:2016aoo}, which provided a bulk geometric condition for good bulk-boundary causality to hold in asymptotically AdS spacetimes, was given a CFT understanding by looking at the Regge limit of boundary OPEs \cite{Afkhami-Jeddi:2017aa}. It seems reasonable that one could use similar techniques to prove the bulk version of (\ref{eqn:FKBound}).

Finally, it would be surprising if this bound were logically separate from the Quantum Focusing Conjecture~\cite{Bousso:2015mna}. Unlike the QFC and related results, the entropy is conspicuously absent from~\eqref{eqn:FKBound}. The lack of any $\hbar$ factors suggest that~\eqref{eqn:FKBound} is more classical than those other bounds,\footnote{We thank Raphael Bousso for emphasizing this point.} but we leave an exploration of a possible relationship to future work.


\begin{acknowledgments}
We would like to thank C.~Akers, R.~Bousso, V.~Chandrasekaran, D.~Marolf, and A.~Shahbazi-Moghaddam for discussions. The work of AL is supported in part by the Berkeley Center for Theoretical Physics, the Department of Defense (DoD) through the National Defense
Science \& Engineering Graduate Fellowship (NDSEG) Program and the National Science Foundation under Grant No. NSF PHY-1748958.
\end{acknowledgments}

\bibliographystyle{utcaps}
\bibliography{all}

\end{document}